\documentclass[conference]{IEEEtran}
\IEEEoverridecommandlockouts
\usepackage{cite}
\usepackage{amsmath,amssymb,amsfonts}
\usepackage{algorithmic}
\usepackage{graphicx}
\usepackage{textcomp}
\usepackage{xcolor}
\usepackage{mathtools}
\usepackage{siunitx}
\usepackage{physics}
\usepackage{breqn}
\usepackage{circuitikz}
\usepackage{xcolor}
\usepackage{tikz}
\usepackage{lettrine}
\usepackage{tabularx}
\usepackage{booktabs}
\usepackage[ruled, lined, linesnumbered, commentsnumbered, longend]{algorithm2e}
\usepackage{dblfloatfix} 

\DeclareMathOperator*{\argmin}{arg\,min}
\newcommand{\matr}[1]{\mathbf{#1}}

\def\BibTeX{{\rm B\kern-.05em{\sc i\kern-.025em b}\kern-.08em
    T\kern-.1667em\lower.7ex\hbox{E}\kern-.125emX}}

\begin{document}

\newcounter{mysfig}
\counterwithin{mysfig}{figure}
\renewcommand\themysfig{(\alph{mysfig})}
\makeatletter
\newcommand\Scaption[1]{%
\refstepcounter{mysfig}%
\vskip.5\abovecaptionskip
  \sbox\@tempboxa{\small\themysfig~#1}%
  \ifdim \wd\@tempboxa >\hsize
    \small\themysfig~#1\par
  \else
    \global \@minipagefalse
    \hb@xt@\hsize{\hfil\box\@tempboxa\hfil}%
  \fi
  \vskip\belowcaptionskip}
\makeatother

\newcolumntype{Y}{>{\centering\arraybackslash}X}

\renewcommand{\Re}{\operatorname{Re}}
\renewcommand{\Im}{\operatorname{Im}}
\newcommand{\Wtilde}[1]{\stackrel{\sim}{\smash{{#1}}\rule{0pt}{1.1ex}}}
\newcommand{\wt}[1]{\widetilde{#1}}

\DeclareRobustCommand\full  {\tikz[baseline=-0.6ex]\draw[thick] (0,0)--(0.3,0);}
\DeclareRobustCommand\dotted{\tikz[baseline=-0.6ex]\draw[thick,dotted] (0,0)--(0.3,0);}
\DeclareRobustCommand\dashed{\tikz[baseline=-0.6ex]\draw[thick,dashed] (0,0)--(0.3,0);}
\DeclareRobustCommand\chain {\tikz[baseline=-0.6ex]\draw[thick,dash dot dot] (0,0)--(0.3,0);}

\definecolor{C0}{HTML}{1f77b4}
\definecolor{C1}{HTML}{ff7f0e}
\definecolor{C2}{HTML}{2ca02c}
\definecolor{C3}{HTML}{d62728}
\definecolor{C4}{HTML}{9467bd}

\title{Low Overhead RF Impedance Measurement by Using Periodic Structures
\thanks{
Manuscript received 15 December 2022; revised 2 February 2023; accepted 06 March 2023.
This work was supported in part by Semiconductor Research Corporation (SRC). {\it (Corresponding author: Muslum Emir Avci.)}
\par The authors are with the Arizona State University, School of Electrical, Computer, and Energy Engineering, Tempe, AZ 85281, USA (e-mail: mavci@asu.edu; sule.ozev@asu.edu).
}
}

\author{
\IEEEauthorblockN{Muslum Emir Avci and Sule Ozev}
}

\newcommand\copyrighttext{%
  \footnotesize \textcopyright 2023 IEEE.  Personal use of this material is permitted.  Permission from IEEE must be obtained for all other uses, in any current or future media, including reprinting/republishing this material for advertising or promotional purposes, creating new collective works, for resale or redistribution to servers or lists, or reuse of any copyrighted component of this work in other works.
  DOI: 10.1109/TMTT.2023.3257641 %
  }
\newcommand\copyrightnotice{%
\begin{tikzpicture}[remember picture,overlay]
\node[anchor=south,yshift=10pt] at (current page.south) {{\parbox{\dimexpr\textwidth-\fboxsep-\fboxrule\relax}{\copyrighttext}}};
\end{tikzpicture}%
}

\maketitle
\copyrightnotice

\begin{abstract}
Due to the need for higher reliability and performance from RF circuits, multi-port reflectometers are increasingly used as low-overhead impedance monitors. 
In this work, using periodic structures as multi-ports is proposed.
Periodic structures impose a new constraint on the multi-port theory and simplify it significantly.
This simplification leads to closed form solution for calibration and measurement procedures. 
The closed-form solution also shows that any arbitrary periodic structure will always have a unique solution for both procedures. 
Therefore, the proposed technique does not rely on frequency-dependent behavior of devices, such as directivity and phase shift to measure impedance. 
This fact leads to increased bandwidth and simplified design procedure.
In addition, proposed multi-port structures can be calibrated by using fewer known loads than existing multi-port techniques.
This fact, coupled with the closed-form solution, reduces computation overhead and test time.
The theory and its robustness against non-idealities, such as part-to-part variation, are verified with Monte-Carlo simulations.
A practical embodiment of the technique is demonstrated with EM simulations and hardware experiments.
In this embodiment, the multi-port structure is embedded into an LC matching network.
Hardware experiments show that the embedded multi-port structure can measure test loads with high accuracy from 1.5 GHz to 3.5 GHz, without degrading matching network performance.

\end{abstract}


\section{Introduction}
\lettrine[lines=2]{I}{} MPEDANCE measurement and monitoring is readily deployed on mobile phone RF front ends, and they are combined with antenna tuners to boost the signal reception capabilities of the phones \cite{balteanu2019rf, boyle2013gain}. 
These sensors also can be used in self-healing circuits to improve their yield and performance under process, voltage, and temperature variations \cite{singh2021low}.
The multi-port technique is a natural candidate for impedance measurement and monitoring due to its low overhead and ease of implementation for Built-in Self-Test (BIST) applications  \cite{solomko2019rf, staszek2021balanced}. 
The multi-port technique can measure complex reflection coefficients ($\Gamma$), thus impedance, by just using scalar measurements provided by the power detectors. 
It is one of the few ways to measure complex impedance without using a tuned receiver. 
This technique generally requires a careful design of the testing structure to produce a unique solution from power detector measurements.
It also requires the use of nonlinear solvers during calibration, and depending on the calibration procedure, the measurement phase. 
This also generates additional issues for convergence, computational complexity, and resources needed for carrying out calibrations and measurements in a timely manner.

Calibration methods in the current literature make different trade-offs between calibration complexity, measurement complexity, the number of constraints, and the number of known loads.
Nonlinear solvers can be used for both calibration and measurement phases to directly solve for network parameters with four known loads \cite{hoer_calibration}.
By using more than four known loads, this direct approach can be used for calibrating  nonidealities in the system, such as power detector nonlinearities \cite{haddadi2012formulation}.
In the cases where these nonidealities are not significant, nonlinear solvers are still necessary to solve for expressions involving exponentiation and multiplication of underlying network parameters.
These expressions can be treated as additional unknowns to simplify calibration and measurement procedures at the expense of using more calibration loads.
In \cite{hunter1985explicit, li1982calibration, ghannouchi1988alternative}, five known loads are used to produce an explicit calibration and measurement procedure. 
The number of known quantities is relaxed in \cite{engen1978calibrating, wiedmann1999new} where calibration reflection coefficients do not need to be known but their magnitudes should be equal.
Another calibration approach is proposed in \cite{staszek2018six},  which requires a known load in addition to several unknown loads which cover the desired measurement region in the Smith Chart.
However, in \cite{engen1978calibrating, wiedmann1999new, staszek2018six} a nonlinear solver (or an optimizer) is required for calibration. 
The physical model of the transmission line and the assumption of high impedance power probes are used to simplify underlying multi-port relations and derive explicit measurement and calibration procedures in \cite{caldecott1973generalized}. 
A comparison of the calibration methods is given in Table-\ref{tab:comparison_calibration}.

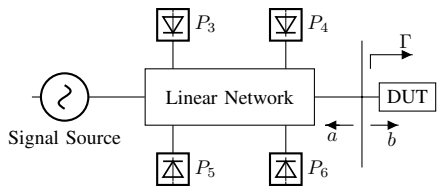
\begin{figure} [t]
    \begin{center}
    \begin{circuitikz}[scale=0.75, transform shape]
    \ctikzset {bipoles/detector/width=.4}
    \draw
        (0.5, 1.5) to[detector, anchor=east, l=$P_3$]++(0, -1)
        (2.5, 1.5) to[detector, anchor=east, l=$P_4$]++(0, -1)
        (0.5, -1.5) to[detector, anchor=east, l_=$P_5$]++(0, 1)
        (2.5, -1.5) to[detector, anchor=east, l_=$P_6$]++(0, 1)
        (4.15, 0) to[short] (3 , 0)
        (3.45, -0.5)  to[short, i=$a$] ++ (-0.25, 0) to[short] ++(0.5, 0)
        (4.00, -0.5)  to[short] ++(0.25, 0) to[short, i_=$b$] ++ (0.25, 0)
        (3.85, -1.25) to[short] ++(0, 2.25) 
        (4, 0.50) to[short] (4, 0.75) to[short] ++(0.5, 0) to[short, i=$\Gamma$] ++ (0.25, 0)
        (-2, 0) to[sV, l_=Signal Source, anchor=east] ++(2, 0);
        \node[draw,minimum width=3cm,minimum height=1cm,anchor=west] at (0, 0){Linear Network};
        \node[draw,minimum width=1cm,minimum height=0.5cm,anchor=west] at (4.15, 0) {DUT};
    \end{circuitikz}
    \end{center}
    \caption{Basic Reflectometer for Measuring $\Gamma$}
    \label{fig:spr_generic}
\end{figure}

In multi-port reflectometers, two ports are used as RF input and DUT ports and the remaining ports are used as measurement ports, which are generally terminated with power detectors.
The number of ports can be adjusted to improve the accuracy and the bandwidth of multi-ports \cite{li1982calibration, caldecott1973generalized}. 
However, the most common implementation of multi-ports is referred to as six-port reflectometers (SPR) where four power detectors are used \cite{engen1977improved}.
These ports are connected by a linear network upon which power detectors are strategically placed to produce a unique solution for the reflection coefficient (Fig. \ref{fig:spr_generic}). 
The design of the linear network determines the multi-port parameters, which in return determines the sensitivity, noise immunity, and the uniqueness of the measurements. A detailed explanation of the parameter selection is provided in \cite{avci2020design}.
For a set of ill-conditioned parameters, multi-port measurements can produce multiple possible reflection coefficient values.
In the current literature, SPRs are generally constructed by a combination of phase shifters, various couplers, and other microwave structures. 
SPRs can be implemented in both integrated circuit level \cite{engen1977improved, lasri2006fully, qayyum201894, wiedmann1997new, hur2015automated}, and board level \cite{avci2020design, singh2021low, staszek2021balanced, donahue2020power}. 
In \cite{donahue2020power}, a multi-port is directly embedded into a distributed matching network constructed with open, and short stubs.

Additional methods that are not based on multi-ports exist for directly or indirectly measuring reflection coefficients. 
These methods generally rely on a directional coupler \cite{grebennikov2013rf, song2009cmos, ji2015novel} or a circulator \cite{paul2011adaptively}, which enables the measurement of the incident and reflected voltages directly. 
Another common technique is the insertion of a series element, such as an inductor \cite{boyle2013gain, yoon20142, van2009adaptive, nicolas2017fully} or a transmission line \cite{avci2022fast} to measure the loading of the circuit. 
These methods generally rely on the change of voltage across the series element terminals since it is related to the loading of the circuit. These methods rely on a known physical model of the element to measure the voltage. 
Such an assumption is not always valid; for example, in RF and mmWave frequencies an inductor is not a simple series element, but it behaves like a two port network with series and shunt components \cite{gil2003simple}.

This work introduces a generalized method for calibrating multi-port reflectometers built with periodic structures and using them for impedance measurements.
The periodic structures introduce a new constraint that simplifies the multi-port theory and leads to an explicit calibration and measurement procedure.
Unlike the existing calibration procedures which require at least five loads and various constraints on the load for explicit solution, the proposed method can use three to four known loads for calibration. 
In addition to a reduced number of calibration loads, developed theory indicates periodic structures will always have well conditioned parameters.
This means that as long as the underlying structure can be modeled as a periodic-structure, the underlying network does not need to be tuned to obtain well-conditioned multi-port parameters. 
Therefore, the proposed technique can be implemented using lumped components and does not require phase shifters or couplers to achieve well-conditioned parameters.
For example, a periodic structure based multi-port can be built by simply partitioning existing lumped components into four segments and placing power detectors at each segment. 
This independence from parameter choice also leads to increased bandwidth of operation and simplifies design procedure.
In summary, advantages of the proposed method can be enumerated as:
\begin{enumerate}
    \item Measurement and calibration procedures are formulated in closed form and the use of nonlinear solvers is not necessary
    \item The proposed method does not require the use of phase shifters and couplers, and can be constructed with lumped components
    \item The proposed calibration method requires 3 loads, less than the existing methods
\end{enumerate}
The proposed technique can be used in a wide variety of metrology applications where multi-port techniques are already used. 
However, the main target application in this paper is BIST for RF circuits. 

The theory, and the measurement and calibration procedures for periodic-structure multi-ports are developed in Section-\ref{sec:methodology}.
The robustness of the method against non-idealities that can be encountered in practice is demonstrated with Monte-Carlo simulations in Section-\ref{sec:sensitivity}.
An experimental circuit, where a periodic-structure multi-port is embedded into a two section LC matching network is designed and verified with EM simulations in Section-\ref{sec:design}.
This design is fabricated and used in hardware experiments. 
The results from these experiments are presented in Section-\ref{sec:experiment}.

\begin{table}[t]
\caption{Comparison of Calibration Methods}
\label{tab:comparison_calibration}
\begin{tabularx}{\linewidth}{@{}YYYYYY@{}}
\midrule
Ref                                          & No. Detectors & Known/Total Cal. Loads & Cal. Procedure & Meas. Procedure \\ \midrule
\cite{hoer_calibration} & 4 & 4/4 & Nonl. solv. & Nonl. solv. \\ \midrule
\cite{haddadi2012formulation}$^{\mathrm{1}}$ & 4 & $\geq7$ & Nonl. solv.   & Nonl. solv. \\ \midrule
\cite{hunter1985explicit, ghannouchi1988alternative} & 4 & 5/5 & Explicit & Explicit \\ \midrule
\cite{li1982calibration}$^{\mathrm{2}}$ & 4 & 4/4 & Explicit & Explicit \\ \midrule
\cite{engen1978calibrating, wiedmann1999new}$^{\mathrm{3}}$ & 4 & 3/5 & Nonl. solv. & Explicit \\ \midrule
\cite{staszek2018six} & 4 & 1/7 & Nonl. solv.   & Explicit \\ \midrule
\cite{caldecott1973generalized}$^{\mathrm{4}}$ & 3 & - & - & Explicit \\ \midrule
\textbf{This work} & 5 & 3/3 & Explicit & Explicit \\ \midrule
\multicolumn{5}{p{\dimexpr\linewidth-2\tabcolsep-2\arrayrulewidth}}{$^{\mathrm{1}}$ The number of calibration loads depends on the specifics of the application. This work also calibrates the power detector non-idealities using cal. loads. } \\
\multicolumn{5}{l}{$^{\mathrm{2}}$ $|\Gamma|\approx 1$ for cal. loads } \\
\multicolumn{5}{l}{$^{\mathrm{3}}$ $|\Gamma|$ should be equal for all cal. loads } \\
\multicolumn{5}{p{\dimexpr\linewidth-2\tabcolsep-2\arrayrulewidth}}{$^{\mathrm{4}}$ Requires high impedance detectors equally spaced on a transmission line  with known $Z_0$. The measurement plane is set and cannot be moved. If an arbitrary measurement plane is used, at least 3 calibration loads are required.}
\end{tabularx}
\end{table}

\begin{figure*}[t]
    \centering
    \includegraphics[width=\linewidth]{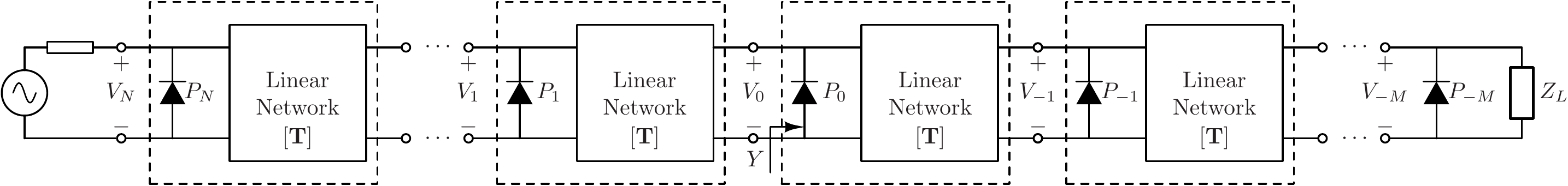}
    \caption{Proposed Measurement System}
    \label{fig:ps_general_form}
\end{figure*}
\section{Periodic Structures as Multi-Ports} \label{sec:methodology}

\begin{figure}[htbp]
    \centering
    \includegraphics[width=0.63\linewidth]{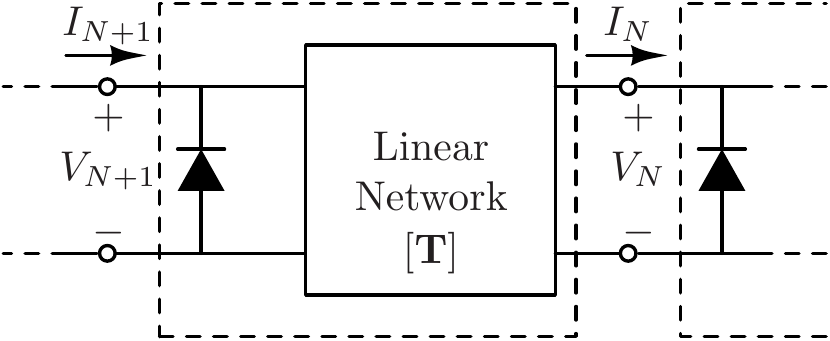}
    \caption{Unit cell of a Periodic Structure}
    \label{fig:unit_cell}
\end{figure}

Multi-ports are generally analyzed and modeled with scattering parameters.
However, due to the repeating nature of the proposed structure, using ABCD/transmission parameters and measuring loads in the admittance domain (at least initially) simplifies the analysis.

Voltages and the currents at the ports of the unit cell (Fig. \ref{fig:unit_cell}) are related through the unit cell's transmission matrix. This relation can be described as:
\begin{equation} \label{eq:t_params}
    \left[\begin{matrix}V_{N+1}\\I_{N+1}\end{matrix}\right] = \left[\begin{matrix}A & B\\C & D\end{matrix}\right] 
    \left[\begin{matrix}V_{N}\\I_{N}\end{matrix}\right].
\end{equation}
Let $\matr{T}$ denote the transmission matrix of the unit cell:  
\begin{equation} \label{eq:T}
    \matr{T} = \left[\begin{matrix}A & B\\C & D\end{matrix}\right].
\end{equation}
The voltage and current at any port in the periodic structure (Fig. \ref{fig:ps_general_form}) can be found by using the cascading property of the transmission matrices, as the following:
\begin{align} \label{eq:R_cascade_relation}
    \left[\begin{matrix}V_{N}\\I_{N}\end{matrix}\right] &= \matr{T}^{N}\left[\begin{matrix}V_{0}\\I_{0}\end{matrix}\right].
\end{align}
The relation between $I_{0}$ and $V_{0}$ in terms of the measured admittance $Y = Z^{-1}$ can be expressed as:
\begin{equation} \label{eq:IVY_relation}
    I_{0} = V_{0} Y.
\end{equation}
$I_{N}$ can be expressed in terms of $V_{N}$ and $Y$ by multiplying both sides of (\ref{eq:R_cascade_relation}) by $1/V_{0}$ and using (\ref{eq:IVY_relation}) as: 
\begin{align} 
   \left[\begin{matrix}V_{N} / V_{0}\\ I_{N} / V_{0} \end{matrix}\right] = \matr{T}^{N} \left[\begin{matrix} 1 \\ Y \end{matrix}\right]. \label{eq:meas_eqn}
\end{align}
Observe that $N$ can be a positive or negative integer. 
This observation will be used to simplify the formulation and derive explicit calibration and measurement procedures.

\begin{figure*}[t!]
    \centering
    \includegraphics[width=\linewidth]{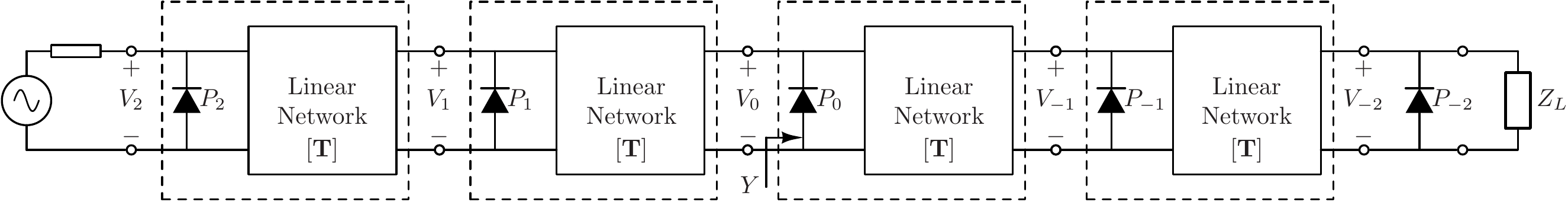}
    \caption{Five detector implementation of the proposed method}
    \label{fig:ps_5detectors}
\end{figure*}

\subsection{Reciprocal Cells with Diagonalizable Matrices}
The transmission matrix can be used for modeling all two-port linear circuits.
This work will concentrate on linear circuits that are reciprocal and have a diagonalizable transmission matrix ($\matr{T}$). 
In RF frequencies, most components, such as inductors, belong to this category. 
If an element has a series and a shunt component, it likely has a diagonalizable transmission matrix (see Appendix for more details). 

If matrix $\matr{T}$ is diagonalizable, it can be written as: 
\begin{align}
\matr{T}^N = \matr{Q} \matr{\Lambda}^N \matr{Q}^{-1}.
\end{align}
For any 2x2 matrix, $\matr{Q}$ can be written as:
\begin{align}
    \matr{Q} = \left[ \begin{matrix} v_1 & v_2 \\ 1 & 1 \end{matrix} \right]
\end{align}
where $\matr{v}_{1,2}=\left[\begin{matrix}v_{1,2} & 1\end{matrix}\right]^T$ are eigenvectors of $\matr{T}$.
Similarly, $\matr{\Lambda}$ can be written as:
\begin{align}
    \matr{\Lambda} = \left[ \begin{matrix} \lambda_1 & 0 \\ 0 & \lambda_2 \end{matrix} \right]
\end{align}
where $\lambda_{1,2}$ are eigenvalues of $\matr{T}$.
For reciprocal circuits $\det\matr{T}=\lambda_1\lambda_2=1$. 
Let $\lambda=\lambda_1$, then the matrix can be expressed as:
\begin{align} \label{eq:T_of_symmetrical_reciprocal_network}
    \matr{T}^N = \frac{1}{v_1 - v_2} 
    \left[ \begin{matrix} v_1 & v_2 \\ 1 & 1 \end{matrix} \right]
    \left[ \begin{matrix} \lambda^N & 0 \\ 0 & \lambda^{-N} \end{matrix} \right]
    \left[ \begin{matrix} 1 & -v_2 \\ -1 & v_1 \end{matrix} \right].
\end{align}
Putting (\ref{eq:T_of_symmetrical_reciprocal_network}) in (\ref{eq:meas_eqn}); normalized voltages are expressed in terms of the eigenvalues, eigenvectors, and load admittance as:
\begin{align}
\frac{V_N}{V_0} = \frac{\lambda^{N} \left(- Y v_{1} v_{2} + v_{1}\right) + \left(Y v_{1} v_{2} - v_{2}\right) \lambda^{-N}}{v_{1} - v_{2}}.
\end{align}
Let $v_2 = kv_1 = kv$, then the normalized voltage becomes:
\begin{align} 
\frac{V_N}{V_0} = \frac{\lambda^{N} \left(Yv k - 1\right) + k \left(1 - Yv\right) \lambda^{-N}}{k - 1}.
\end{align}
Let $w = (Yvk - 1)/(k-1) - 1/2$, then the normalized voltage can be expressed as:
\begin{align} \label{eq:Vn_eigenY}
\frac{V_N}{V_0} = \lambda^{N} \left(\frac{1}{2} + w \right) + \lambda^{-N} \left(\frac{1}{2} - w\right).
\end{align}
From (\ref{eq:Vn_eigenY}), one can observe that, each introduced measurement port is linearly independent and does not increase the number of unknowns. 
This central finding is key to reducing the number of the calibration loads and computational complexity. 
Note that the load of interest $Y$ is a bilinear transform (Möbius transform) of the $w$, therefore after $w$ is measured, $Y$ can be solved in terms of $w$.
This will be investigated in the calibration section.

\subsection{Power Detector Measurements}
To solve for $w$, power detector/voltage measurements can be used.
It is assumed that $\lambda$ is known and found by calibration.
After $w$ is found, an error-box correction approach can be used to find the load value explicitly. 
Error-box parameters also will be found during the calibration phase. 

The output of each power detector can be formulated as:
\begin{align} \label{eq:PN}
    P_N = g \abs{V_N}^2
\end{align}
where $g$ is the gain of the power detector. 
Power measurements will be normalized by a predetermined "reference" power measurement to cancel out the detector gains and the scaling effects of the test signal amplitude.
Therefore, the measurement ratio ($M_N$) is defined as:
\begin{align} \label{eq:Mn_def}
    M_N = \frac{P_N}{P_0} = \abs{ \frac{V_N}{V_0} }^2
\end{align}
where $V_0$ is the reference voltage measurement.
To find an explicit formulation, (\ref{eq:Vn_eigenY}) needs to be rearranged as the following:
\begin{align} \label{eq:Mn_def2}
    M_N = \abs{\frac{V_N}{V_0}}^2 = \abs{(\lambda^{N} - \lambda^{-N}) w + \frac{1}{2} (\lambda^N + \lambda^{-N})}^2.
\end{align}
Let $J_N=\lambda^{N} - \lambda^{-N}$ and $L_N=\lambda^{N} + \lambda^{-N}$, then $M_N$ can be expressed as:
\begin{align} \label{eq:MN}
    M_N &= |J_N|^2|w|^2 + \frac{1}{4} |L_N|^2 + \Re{\{J_N L_N^*w\}}
\end{align}
and $M_{-N}$ can be expressed as:
\begin{align} \label{eq:MN_neg}
    M_{-N} &= |J_N|^2|w|^2 + \frac{1}{4} |L_N|^2 - \Re{\{J_N L_N^*w\}}.
\end{align}
This symmetry can be used to separate the terms as follows:
\begin{align}
    A_N = M_N + M_{-N} &= 2|J_N|^2|w|^2 + \frac{1}{2} |L_N|^2 \\
    B_N = M_N - M_{-N} &= 2\Re{\{J_N L_N^*w\}}.
\end{align}
These concepts and definitions will be used later in the calibration section as well.
If $\lambda$ (therefore $J_N$ and $L_N$ is known), $\abs{w}$ can be expressed as:
\begin{align}
    \abs{w}^2 = \frac{A_N - 0.5 \abs{L_N}^2}{2\abs{J_N}^2}.
\end{align}
Similarly, $\Re{\{J_NL_N^*w\}}$ can be expressed as:
\begin{align}
    \Re{\{J_N L_N^*w\}} = \frac{B_N}{2}.
\end{align}
Using $\lambda$, $\abs{w}$ and $\Re{\{J_N L_N^*w\}}$, the imaginary part can be expressed as:
\begin{align}
    \Im{\{J_N L_N^*w\}} = \pm \sqrt{|J_N L_N|^2|w|^2 - \Re^2{\{J_N L_N^*w\}}}.
\end{align}
Finally, from the measurements, two possible solutions for $w$ are found:
\begin{align}
    w_{1,2} = \frac{\Re{\{J_N L_N^*w\}} \pm  j\Im{\{J_NL_N^*w\}}}{J_NL^*_N}.
\end{align}
So far, only two measurements are used. 
At least one more measurement (i.e., $N+1$) is necessary to find the correct root. 
Therefore, at least three measurements (four power detectors) are needed to find $w$ unambiguously.
The correct solution can be selected by plugging values back in (\ref{eq:Mn_def2}) and choosing the solution that has the smallest difference between the predicted and actual power measurement. 
This final measurement step can be expressed as:
\begin{align}
    \argmin_{w\in\{w_1, w_2\}} \abs{M_{N+1} - \abs{ \lambda^2 \left( \frac{1}{2} + w \right) - \lambda^{-2} \left( \frac{1}{2} - w \right)  }^2}.
\end{align}
After $w$ is solved, the calibration and system information can be used to extract $Y$, $Z$ or $\Gamma$. 
The given formulations are correct for any $N$. 
Due to its convenience and ease of calculation $N=1$ is selected. Therefore; $M_{-1}$, $M_1$, $M_2$ will be used for calculating $w$. 

\subsection{Calibration}
The calibration procedure will be performed in two steps. 
In the first step, $\lambda$ will be found. 
After $\lambda$ is known, three distinct loads can be used to find the bilinear relationship between $w$ and the load $Z_L$ (Fig. \ref{fig:ps_5detectors}).

The notation introduced before is expanded to accommodate for the different calibration loads.
For load $i$, the measurement expression is given as:
\begin{align}
    M_{Ni} &= |J_N|^2|w_i|^2 + \frac{1}{4} |L_N|^2 + \Re{\{J_N L_N^*w_i\}}.
\end{align}
Similarly, $A_{Ni}, B_{Ni}$ can be expressed as:
\begin{align}
    A_{Ni} = M_{Ni} + M_{-Ni} &= 2|J_N|^2|w_i|^2 + \frac{1}{2} |L_N|^2 \\
    B_{Ni} = M_{Ni} - M_{-Ni} &= 2\Re{\{J_N L_N^*w_i\}}.
\end{align}
For simplicity, assume that there are five power detectors, and they are arranged as shown in Fig. \ref{fig:ps_5detectors}.
Then, for load $i$, there will be four measurements ($M_{-2i}$, $M_{-1i}$, $M_{1i}$, $M_{2i}$).

Initially, measurements for two different loads $i=1,2$ are used for finding $\lambda$, which is expressed in polar coordinates $\lambda=r e^{j\theta}$.
The quantities that can be calculated/inferred from the measurements are denoted as $W_{m}$, to differentiate between known quantities from unknown quantities up to that point.

First, observe that $J_2=J_1 L_1$. Based on this $|L_1|^2$ can be found by: 
\begin{align} \label{eq:W1}
    W_1 &= \frac{A_{2, 1} - A_{2, 2}}{A_{1,1} - A_{1, 2}} \\
    &= |L_1|^2.
\end{align}
which in turn can be expressed in polar coordinates:
\begin{align}
    W_1 = r^2 + r^{-2} + 2 \cos 2\theta.
\end{align}
Using $W_1 = |L_1|^2$, the following can be found:
\begin{align}
    W_1 A_{1,1} - A_{2, 1} &= \frac{ \abs{\lambda - \lambda^{-1}}^4 - \abs{\lambda^2 - \lambda^{-2}}^2 }{2} \\
    &= 2 \cos2\theta (r^2 + r^{-2}) + 2.
\end{align}
Let $W_2$ be:
\begin{align}
    W_2 &= \frac{W_1 A_{1,1} - A_{2, 1} - 2}{2} \\
    &= \cos2\theta (r^2 + r^{-2})
\end{align}
by using $W_1$ and $W_2$, $\cos2\theta$ can be found by solving:
\begin{align}
    0 = 2 \cos^2 2\theta - W_1 \cos2\theta + W_2.
\end{align}
There are two roots in this equation. 
One of the roots corresponds to $\cos2\theta$, and the other root corresponds to $(r^2 + r^{-2})/2$. 
This fact also can be verified by observing that the product of the roots of this equation is equal to $W_2/2=\cos2\theta (r^2 + r^{-2})/2$.
Luckily, the roots can be assigned to the corresponding expressions unambiguously by observing that $(r^2 + r^{-2})/2 \geq 1$ and $\cos2\theta \leq 1$, as the following:

\begin{align}
    \cos2\theta &= \frac{W_1 - \sqrt{W_1^2 - 8 W_2}}{4} \\
    W_3 = r^2 + r^{-2} &= \frac{W_1 + \sqrt{W_1^2 - 8 W_2}}{2}. \label{eq:r2_rm2}
\end{align}
Using $W_3$, (\ref{eq:r2_rm2}) can be succinctly expressed as:
\begin{align}
    0 = r^4 - W_3 r^2 + 1.
\end{align}
By solving this equation, all possible solutions for the $r$ are found:
\begin{align}
     r = \pm\sqrt{\frac{W_3 \pm \sqrt{W_3^2 - 4}}{2}}.
\end{align}
As expected, there are four roots for this equation and they correspond to $\pm r, \pm 1/r$. 
The sign of the $r$ does not matter, because only $r^2$ term is used in the measurements.
$1/r$ is also a valid solution since there are two eigenvalues and they are inverse of each other. 
Both eigenvalues will produce valid and correct measurements. 

The only term that cannot be solved unambiguously is the imaginary part of $\lambda$. 
To solve it unambiguously, the sign of $\sin2\theta=\pm \sqrt{1-\cos^2 2\theta}$ needs to be determined, which is not possible with the given calibration procedure. 
In practice, this is not a big concern since the sign of this expression is related to the intrinsic properties of the unit cell. Therefore, the sign can be determined {\it a priori}. 
In the cases where this is not feasible, a fourth load can be used as a test load to determine the sign. 
If the fourth load measurement does not correspond to the known value, the imaginary sign is changed, and measurement is performed again.
This completes the first step of the two-step calibration.

\begin{figure}[ht!]
    \begin{minipage} {\linewidth}
        \centering
        \includegraphics[width=\textwidth]{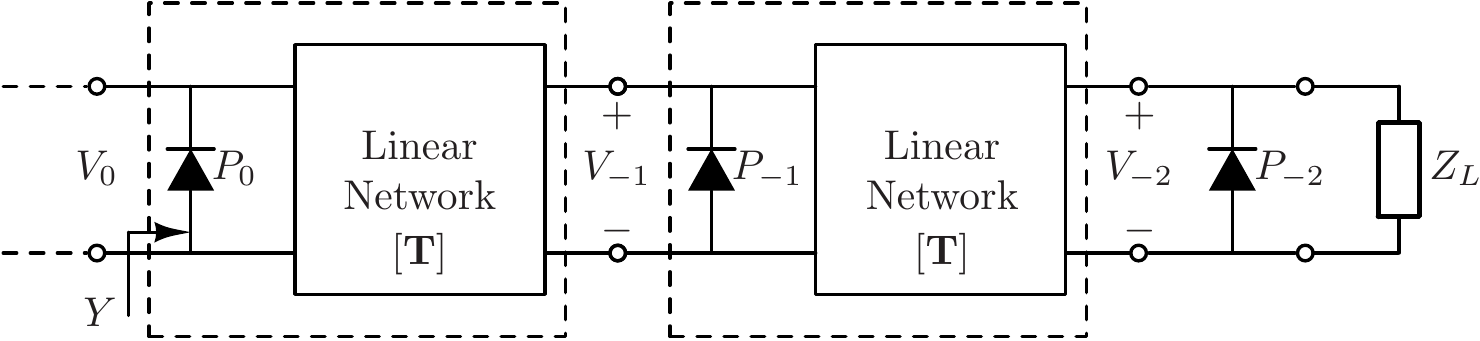}
        \Scaption{Measurement Plane}
        \label{fig:measurement_plane}
        \vspace{3mm}
    \end{minipage}
    \begin{minipage} {\linewidth}
        \centering
        \includegraphics[width=0.4\textwidth]{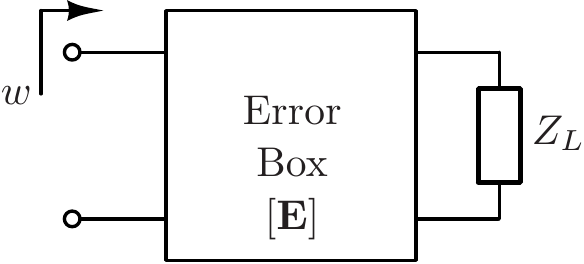}
        \Scaption{Error Box Equivalent}
        \label{fig:error_box}
        \vspace{3mm}
    \end{minipage}
    \caption{Transformation Steps Demonstrating $Y \mapsto w \mapsto Z_L$}
    \label{fig:error_box_calibration}
\end{figure}

In the second step, error-box parameters that relate $w$ to $Z_L$ will be solved (Fig. \ref{fig:error_box_calibration}). 
Now, that $\lambda$ is solved, $w_i$ for each load can be measured as described in the previous section. 
Each $w_i$ is a bilinear transform of $Y_i$, which in return is a bilinear transform of the load of interest $Z_{Li}$.
The composition of bilinear transforms is a bilinear transform. 
Therefore, there exists a bilinear relationship between $w_i$ and $Z_{Li}$, which can be expressed as:
\begin{align}
    Z_{Li} = \frac{a w_i + b}{c w_i + 1}.
\end{align}
Using three distinct known loads, $a, b$ and $c$ are determined, which are the terms for the error box. 
Since the impedance, admittance, and reflection coefficient are bilinear transforms of each other, error-box parameters can be solved for any of these measurement targets.

To use the outlined calibration method, two pairs of measurement ratios ($M_{\pm N}$) are needed, which translate to five power detectors. 
This calibration procedure does not require any nonlinear solvers. 
However, obtaining an initial solution with this method and using a nonlinear solver to solve for eigenvalues can provide accuracy benefits. 
In the experiments and sensitivity simulations, closed form solution is determined to be sufficiently accurate. 
To maximize the information each load contributes, calibration loads should be selected as far apart on the Smith chart and complex loads should be preferred when possible.

\subsection{Measurement Port Considerations}
\begin{figure} [ht!]
    \centering
    \includegraphics[width=\linewidth]{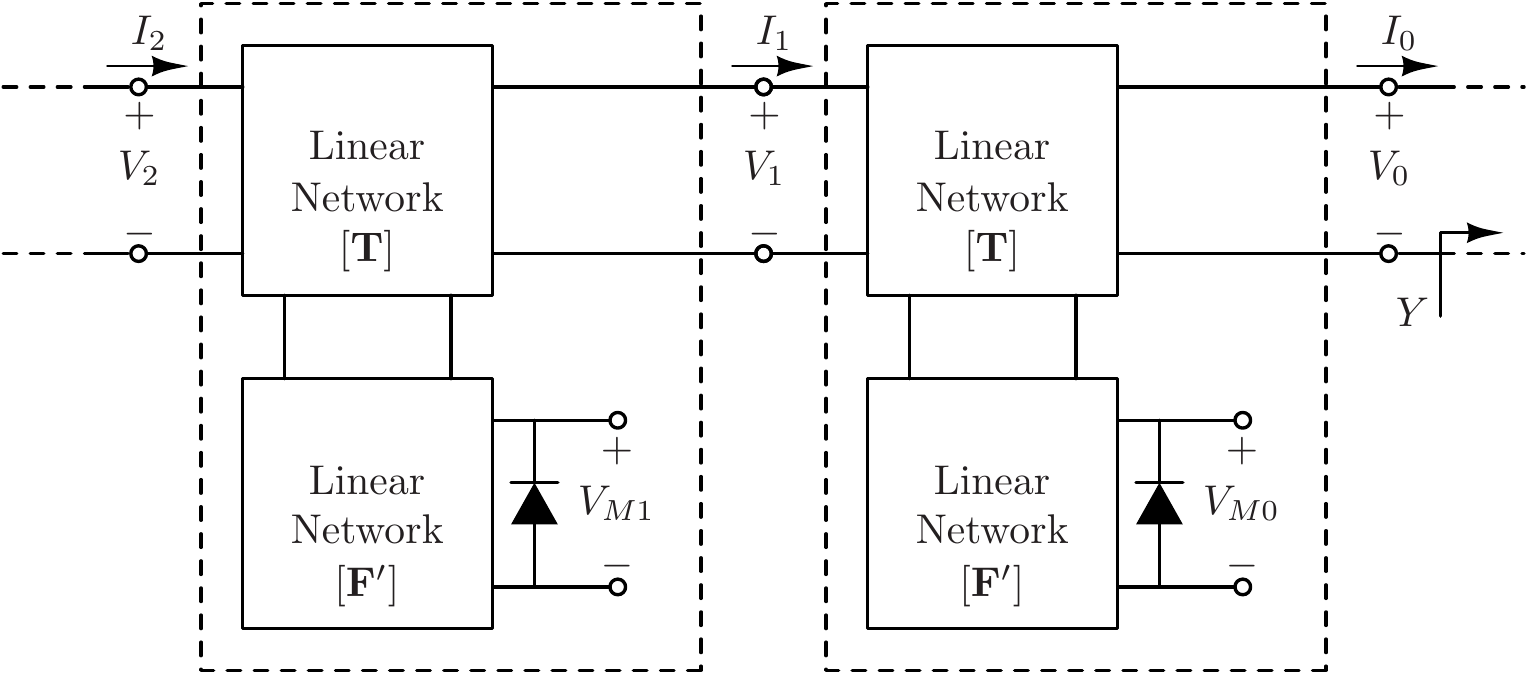}
    \caption{Measurements with Arbitrary Transfer Matrix}
    \label{fig:measurement_port_general}
\end{figure}

So far it has been assumed that voltages can be directly measured on the edges of unit-cells. This may not be practical in some applications. 
In this section, it is demonstrated that power detectors can be placed anywhere in the unit cell, and the outlined method would still work without any modifications.

First, consider the case shown in Fig. \ref{fig:measurement_port_general}. 
One node in the circuit is exposed as a measurement port and the linear network that is connected to the measurement port can be arbitrarily complex.
The following relation can still be expressed between the measured voltage ($V_{M0}$), the input voltage ($V_0$) and the current ($I_0$) to unit cell:
\begin{align}
    V_{M0} = A V_0 + B I_0
\end{align}
where $A$, and $B$ are arbitrary network parameters. Since $V_0$ is the reference port, where the admittance measurement plane is, the current can be expressed as $I_0 = Y V_0$. Then, $V_{M0}$ becomes:
\begin{align} \label{eq:measured_v_relation}
    V_{M0} = V_0 (A + B Y).
\end{align}
This relation can be expressed in matrix form as the following:
\begin{align}
    \left[\begin{matrix}V_{M0} \\ V_{0} Y \end{matrix}\right] &= \left[\begin{matrix} A &  B \\ 0 & 1 \end{matrix}\right] \left[\begin{matrix}V_{0} \\ V_0 Y \end{matrix}\right].
\end{align}
Let $\matr{F^{\prime}}$ denote this transformation matrix
\begin{align}
    \matr{F^{\prime}} = \left[\begin{matrix} A &  B \\ 0 & 1 \end{matrix}\right].
\end{align}
This matrix is invertible, given $A\neq0$. 
Using $\matr{F^{\prime-1}}$ and $\matr{T}$, the voltage and the current at the next port is found, then $\matr{F^{\prime}}$ is applied again to find measured voltage at the next port as
\begin{equation}
     \left[\begin{matrix}V_{M1} \\ I_1 \end{matrix}\right] = \matr{F^{\prime}} \matr{T} \matr{F^{\prime-1}} \left[\begin{matrix}V_{M0} \\ V_{0} Y \end{matrix}\right] 
\end{equation}
both sides are divided by the reference voltage, $V_{M0}$,
\begin{equation}
     \left[\begin{matrix}V_{M1} / V_{M0} \\ I_1 / V_{M0} \end{matrix}\right] = \matr{F^{\prime}} \matr{T} \matr{F^{\prime-1}} \left[\begin{matrix} 1 \\ V_{0} / V_{M0} Y \end{matrix}\right] 
\end{equation}
and by using (\ref{eq:measured_v_relation}), $V_0$ is written in terms of $V_{M0}$:
\begin{equation}
     \left[\begin{matrix}V_{M1} / V_{M0} \\ I_1 / V_{M0} \end{matrix}\right] = \matr{F^{\prime}} \matr{T} \matr{F^{\prime-1}} \left[\begin{matrix} 1 \\ Y / (A + BY) \end{matrix}\right].
\end{equation}
Let $Y_M = Y / (A + BY)$, then for any arbitrary measurement port the following can be written:
\begin{equation}
     \left[\begin{matrix}V_{MN} / V_{M0} \\ I_N / V_{M0} \end{matrix}\right] = \matr{F^{\prime}} \matr{T}^N \matr{F^{\prime-1}} \left[\begin{matrix}V_{M0} \\ Y_M \end{matrix}\right].
\end{equation}
Observe that $\matr{F^{\prime}} \matr{T}^N \matr{F^{\prime-1}} = (\matr{F^{\prime}} \matr{T} \matr{F^{\prime-1}})^N$, thus $\matr{T'} = \matr{F^{\prime}} \matr{T} \matr{F^{\prime-1}}$ can be used as the unit cell transmission matrix and all previously developed methods can be applied without any modification. 
Also observe that eigenvalues of $\matr{T'}$ and $\matr{T}$ are equal.
The measured quantity $w$, will be a bilinear transformation of $Y_M$, which can solved by using error-box calibration as described previously. 

\begin{figure} [t]
    \centering
    \includegraphics[width=\linewidth]{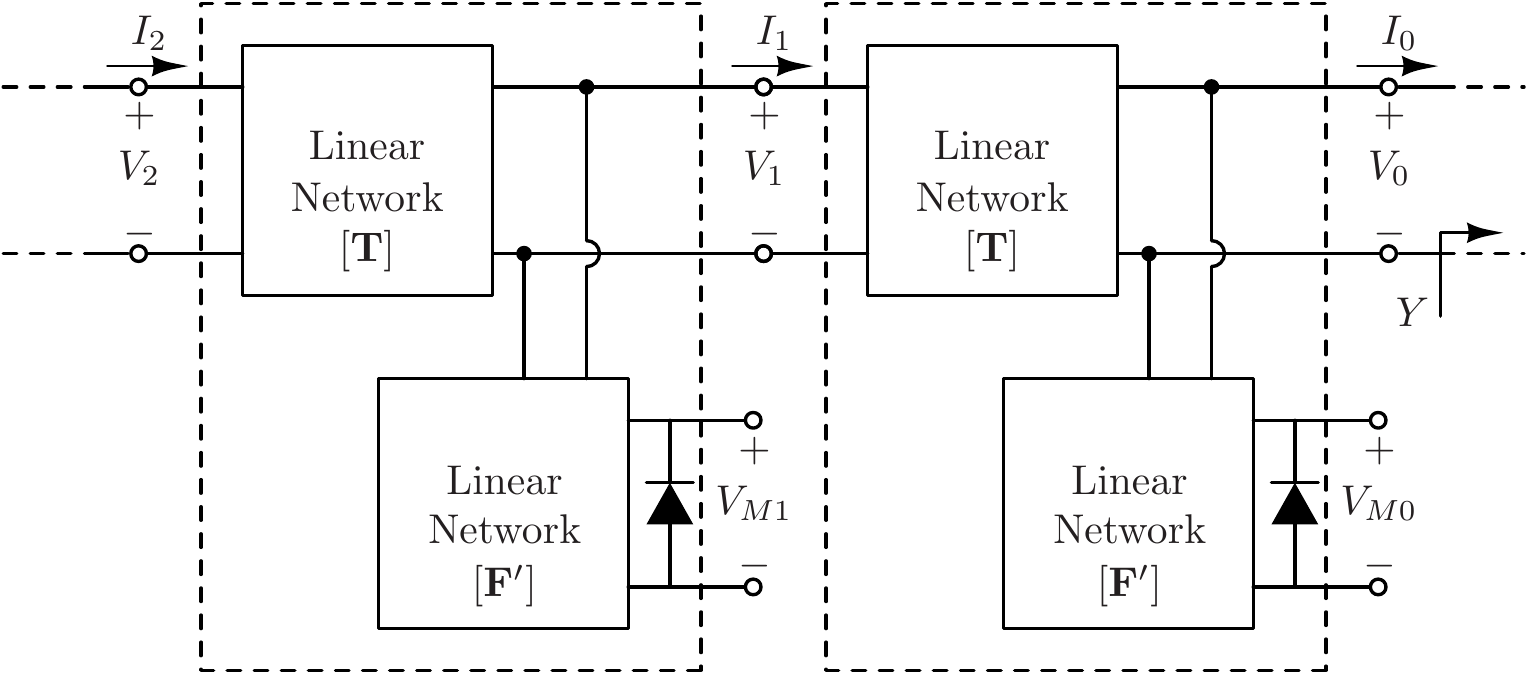}
    \caption{Measurements with Detectors Placed on Unit-Cell Edge}
    \label{fig:measurement_port_edge}
\end{figure}

A special case of measurement port placement occurs when measurement ports are placed on the edges of the repeating unit cells (Fig. \ref{fig:measurement_port_edge}). 
In this configuration, $Y_M$ would be the scaled version of $Y$.
Also, if measurement ports are placed on the edge, one less repeating cell can be used than the arbitrary configuration. Due to these advantages, measurements ports are placed on the edge in the design example.

\subsection{Dynamic Range}
One of the main design goals is to select the dynamic range of the multi-port so that power at each port can be accurately measured and quantified. 
If the dynamic range requirement is too high, detectors may not be able to measure power accurately. 
Similarly, if the dynamic range is too low, quantifying measurements becomes a challenge.

The dynamic range requirement for the power detectors is determined by multi-port parameters and the maximum VSWR or $|\Gamma|_{max}$ that should be measured.
Therefore, to determine the dynamic range requirements, $M_N$ is expressed in terms of $\Gamma$ and multi-port parameters as the following:
\begin{align}
    M_N &= \abs{J_N \frac{a \Gamma + b}{c\Gamma + 1} + \frac{1}{2} L_N}^2
\end{align}
where $a$, $b$, $c$ are error-box terms that relate $\Gamma$ to $w$. These parameters are functions of the eigenvectors, measurement port parameters and the measurement plane. 
$M_N$ can be rearranged as the following to obtain the canonical six-port formula:
\begin{align} \label{eq:Mn_qn}
    M_N = K_N \abs{ \frac{ \Gamma + q_N }{c \Gamma + 1} }^2
\end{align}
where $K_N$ and $q_N$ are defined as the following:
\begin{align}
    K_N &= \abs{(J_{N} a + 0.5 L_{N} c)}^2 \\ 
    q_N &= (J_{N} b + 0.5 L_{N}) / (J_{N} a + 0.5 L_{N} c).
\end{align}
Each port has an unique $q_N$. Let $|q|_{max}$, $|q|_{min}$ be the maximum and the minimum magnitude of $q_N$ respectively.
Also note that if $|c| \approx 0$, then $q_N$ would correspond to the circle centers of the six-ports \cite{engen1977six}. 

It is assumed that the test signal power can be adjusted to accommodate the operating region (linear region) of the power detectors. Therefore, the dynamic range ($DR$) of the power detectors needs to be greater or equal to the ratio of the largest ($P_{max}$) and the smallest value ($P_{min})$ that power detectors can obtain in a single measurement, which can be expressed as:
\begin{align}
    DR = \frac{P_{max}}{P_{min}} = \frac{M_{max}}{M_{min}}
\end{align}
where $M_{max}$, $M_{min}$ are the maximum and minimum values that $M_N$ can have for any $N$ (including zero) in a single measurement. 
$M_{max}$ and $M_{min}$ can be found by maximizing and minimizing the nominator and denominator of (\ref{eq:Mn_qn}) appropriately by using the triangle inequality. This approach gives the upper bound for dynamic range as the following:
\begin{align} \label{eq:dynamic_range}
    DR = \left( \frac{ \abs{q}_{max} + \abs{\Gamma}_{max} } {\abs{q}_{min}  - \abs{\Gamma}_{max}} \cdot \frac{1 + \abs{c} \abs{\Gamma}_{max} } { 1 - \abs{c} \abs{\Gamma}_{max} } \right)^2.
\end{align}
To have a bounded dynamic range, $|q|_{min}$ and $1/|c|$ should be bigger than $|\Gamma|_{max}$. 
The dynamic range of a particular port can be decreased by increasing $|q|_{min}$ or decreasing $|c|$, $|q|_{max}$. 
Therefore, the dynamic range can be adjusted by strategically selecting eigenvalues, eigenvectors, power detector placement and the measurement plane.  
In a more traditional multi-port/six-port $|c|\approx0$ achieved generally by using directional couplers, a similar result can be obtained by placing the power detector inside the unit cell. 
Unfortunately, placing power detectors on the edge of the unit-cell would result in $|c| > 0$, increasing the required dynamic range. 
Another way to decrease the required dynamic range is achieved by increasing the number of detectors. It is unlikely each detector would reach its extreme values at the same time and only one pair of $M_{\pm N}$ needs to be measured accurately to find $\Gamma$. 

\section{Monte-Carlo Simulation \& Sensitivity Analysis} \label{sec:sensitivity}
\begin{figure} [htb!]
    \centering
    \includegraphics[width=0.75\linewidth]{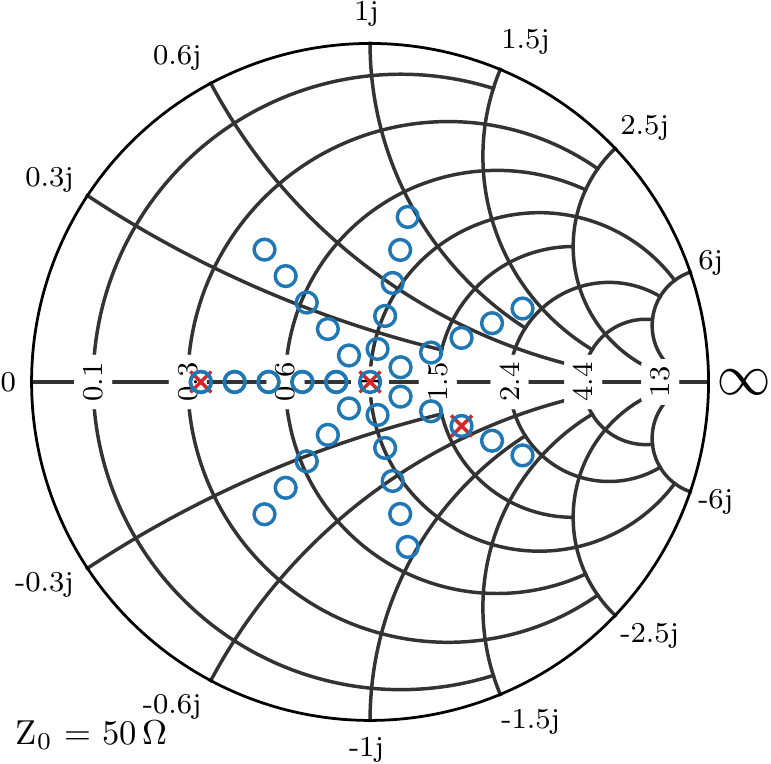}
    \caption{Sensitivity Test Loads, Calibration Loads are shown with ({\color{C3} $\cross$}) }
    \label{fig:sensitivity_smith}
\end{figure}
\begin{figure*} [t!]
    \centering
    \includegraphics[width=\linewidth]{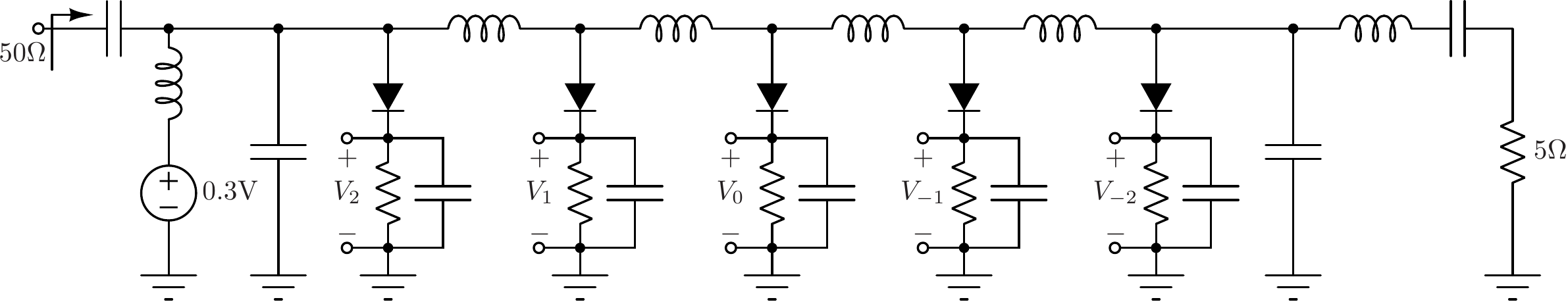}
    \caption{Matching Network with Power Detectors }
    \label{fig:mn_w_pds}
\end{figure*}
\begin{figure}[t!]
    \begin{minipage} {\linewidth}
        \centering
        \includegraphics[width=\textwidth]{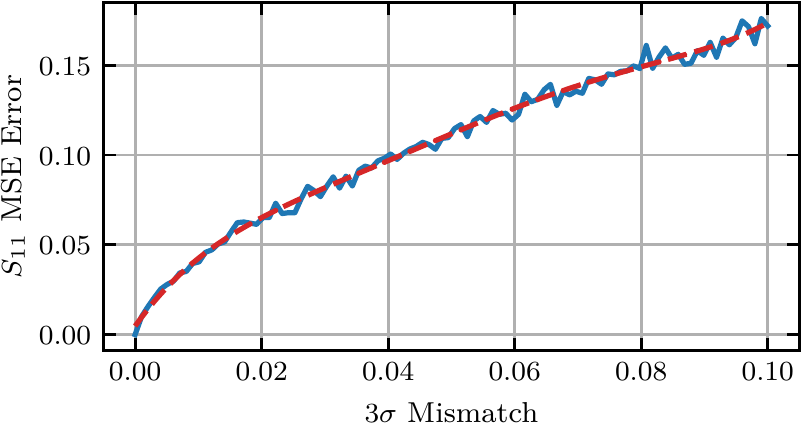}
        \Scaption{MSE between Loads and Measurements in $S_{11}$ Domain}
        \label{fig:combined_mse}
        \vspace{3mm}
    \end{minipage}
    \begin{minipage} {\linewidth}
        \centering
        \includegraphics[width=\textwidth]{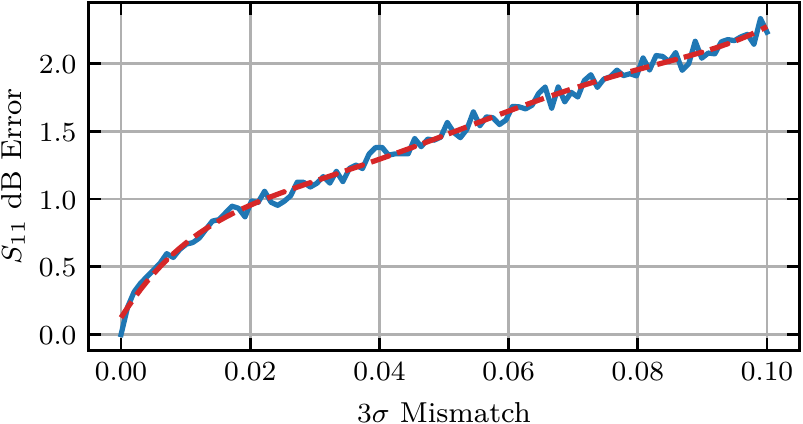}
        \Scaption{dB Error between Loads and Measurement in $S_{11}$ Domain}
        \label{fig:combined_db}
        \vspace{3mm}
    \end{minipage}
    \caption{ Monte-Carlo Simulation, ({\color{C0} \full}) Calculated, ({\color{C3} \dashed}) Polynomial Trend}
    \label{fig:sensitivity_results}
\end{figure}

The main consideration of the practicality of the proposed approach are mismatches between different unit cells and power detectors.
Due to process variations, each unit cell will have slightly different $\matr{T}$, $\matr{F^\prime}$, and power detector gain. 
Therefore, the proposed method needs to be evaluated against such non-idealities to determine its robustness and to determine how tight tolerances should be in an application. 
To achieve this goal, Monte-Carlo simulations are conducted with different $\matr{T}$, $\matr{F^\prime}$, and power detector gain are used. 
In the simulations, power detector voltages, $V_N$, will be measured as:
\begin{align}
    \left[\begin{matrix} V_N \\ I_N \end{matrix}  \right] = \matr{F^\prime} \matr{T}^{N-1} \left[ \begin{matrix} 1 \\ Y \end{matrix} \right]
\end{align}
for $N=1\dots5$, and $Y$ will be selected as uniformly distributed loads across the Smith Chart as shown in Fig. \ref{fig:sensitivity_smith}. The VSWR for the test loads is under 3:1 ($S_{11} < -3$ dB), to simulate a realistic application. For each load, 1000 random mismatches are generated.
$M_N$ defined as:
\begin{align}
    M_N = \abs{\frac{V_N}{V_3}}.
\end{align}

First, two matrices $\matr{T}$ for the unit cell, and $\matr{F^\prime}$ for the measurement port are constructed.
The real and imaginary part of the entries of $\matr{T}$ ($t_{ij}$) and $\matr{F^\prime}$ ($f_{ij}$) will be randomly sampled from the uniform distribution $U[-1, 1]$.

After verifying that the randomly constructed matrices have an inverse, $\matr{T}$ will be normalized such that $\det\matr{T}=1$.
This ensures that $\matr{T}$ represents a physical reciprocal circuit, which is an underlying assumption for the proposed theory.
Now that the matrix for the unit cell and the measurement port are constructed, random mismatches can be added for each unit cell to introduce non-idealities. 
For each unit cell, two mismatch matrices will be constructed, one for the $\matr{T}$ and another one for the $\matr{F^\prime}$. 
These mismatch matrices will be term-wise multiplied with the unit cell matrices.
$\matr{T}$, $\matr{F^\prime}$ with mismatches added is denoted as $\matr{T}_N$, $\matr{F^\prime}_N$ and their entries are denoted as $t_{Nij}$, $f_{Nij}$ respectively.
These entries can be expressed as
\begin{align}
    t_{Nij} = t_{ij} \cdot m^t_{Nij} \\
    f_{Nij} = f_{ij} \cdot m^f_{Nij}
\end{align}
where, $m^t_{Nij}$, $m^f_{Nij}$ are added mismatches to respective entries.
These mismatches will be sampled from the normal distribution with a mean ($\mu$) one. 
\begin{align} \label{eq:matrix_mismatch_dist}
    m^t_{Nij} &\sim \mathcal{N}(1, \sigma_t^2) \\
    m^f_{Nij} &\sim \mathcal{N}(1, \sigma_f^2).
\end{align}
The last non-ideality that will be simulated is gain mismatch between power detectors. 
To model this, measured voltage is multiplied with a randomly sampled gain $g_N \sim \mathcal{N}(1, \sigma_g^2)$. 
The mean of the gain is selected as one since voltages are normalized.
Therefore, the measured voltage at each port can be expressed as:
\begin{align}
    V_N = \left[ \begin{matrix} g_N & 0 \end{matrix} \right] \matr{F^\prime}_N \prod_{i=1}^{N-1} \matr{T}_{i}.
\end{align}
In order to determine the tolerance, the $3\sigma$ of all mismatch distributions is swept from 0 to 0.1. 
This would mean that 99.9\% of units have less than the specified mismatch. For example, if $3\sigma=0.1$ this would mean 99.9\% of mismatches are below 10\%.
All mismatches ($\matr{T}$, $\matr{F^\prime}$, gain) will be simulated together to determine maximum tolerable mismatch.

The randomly created systems will measure $S_{11}$ of the predetermined set of loads that are shown in Fig. \ref{fig:sensitivity_smith}. 
The system is calibrated using the three known $S_{11}$ values by using the method described in the previous section.

Simulation results showed that (Fig. \ref{fig:sensitivity_results}) to keep measurement error under 1 dB, all combined mismatches should be below 2\%.
This is easily achievable by discrete parts, and it is expected to be easily achievable in an integrated application since most RF components are much bigger than the smallest node feature size. 

It is also determined that the proposed technique has highest sensitivity to measurement port errors ($\matr{F^\prime}$), followed by gain mismatches. 
The proposed technique is least sensitive to the mismatches between unit cells ($\matr{T}$).

\section{Design Example} \label{sec:design}

For hardware demonstration, one embodiment of the proposed technique is designed. 
In this design, a periodic-structure multi-port is embedded  into a two-section LC matching network that transforms \SI{5}{\ohm} to \SI{50}{\ohm}.
The \SI{5}{\ohm} represents the impedance seen by a hypothetical PA, and \SI{50}{\ohm} represents the impedance of a hypothetical antenna.
In the simulation and in the experiment, developed method is used to detect changes in the \SI{50}{\ohm} port.

\begin{figure} [htb!]
    \centering
    \includegraphics[width=0.65\linewidth]{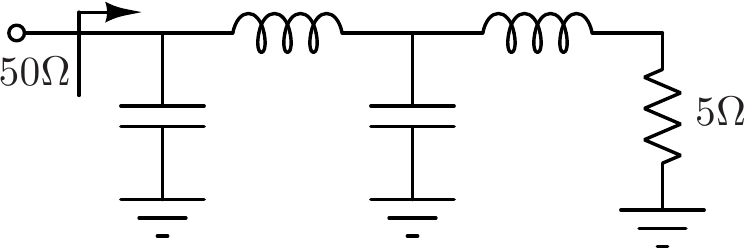}
    \caption{Two Section LC Matching Network}
    \label{fig:mn_wo_pds}
\end{figure}
\begin{figure}[t!]
    \begin{minipage} {\linewidth}
        \centering
        \includegraphics[width=\textwidth]{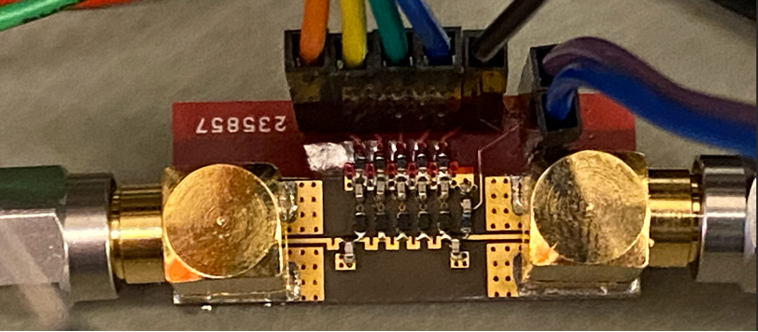}
        \Scaption{Matching Network with Power Detectors}
        \label{fig:mn_w_pds_pcb}
        \vspace{3mm}
    \end{minipage}
    \begin{minipage} {\linewidth}
        \centering
        \includegraphics[width=\textwidth]{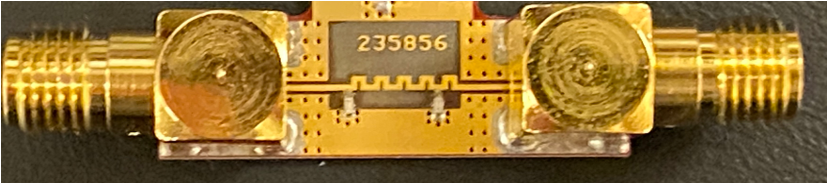}
        \Scaption{Matching Network without Power Detectors}
        \label{fig:mn_wo_pds_pcb}
        \vspace{3mm}
    \end{minipage}
    \caption{Fabricated Boards}
    \label{fig:fabricated_boards}
\end{figure}
 \begin{figure} [b]
    \centering
    \includegraphics[width=0.75\linewidth]{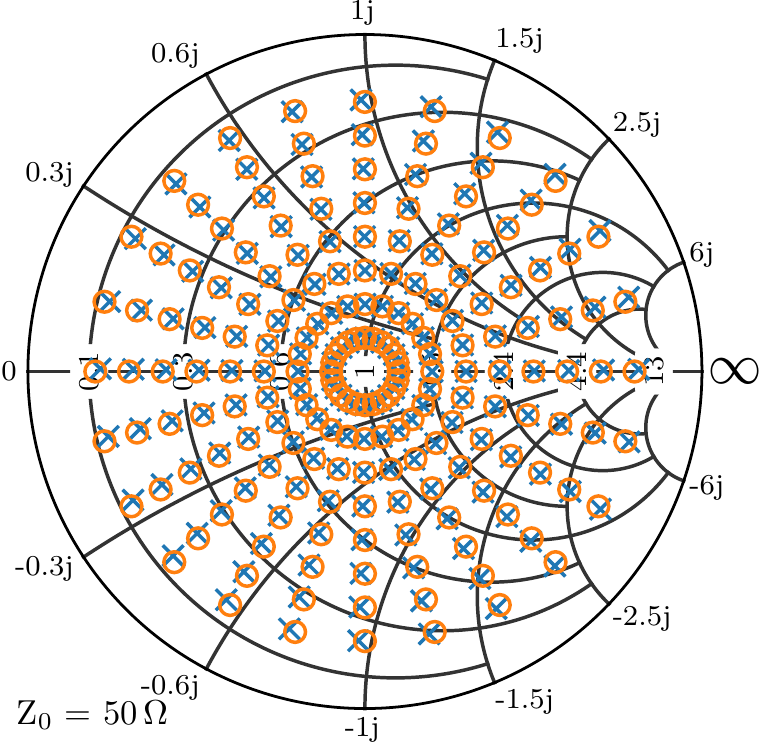}
    \caption{Design Simulation Results - Test Loads ({\color{C1} $\circ$}), Measured Values ({\color{C0} $\cross$})}
    \label{fig:sim_smith_chart}
\end{figure}

The two-section LC matching network without any modifications is shown in Fig. \ref{fig:mn_wo_pds}.
A multi-port based on the proposed measurement method is embedded into this network by dividing the inductor on the left into four smaller inductors (Fig. \ref{fig:mn_w_pds}).
Simple peak detectors are connected to the edges of the divided inductors to measure the voltage incident upon the nodes. 
The peak detector diodes (BAT15-02LRH) are biased such that they are barely open during operation.
To achieve this, high value resistors (\SI{100}{\kilo\ohm}) are connected in series to diodes, so when a diode turns on, the current through them will be minimal. 
Finally, a shunt capacitance is connected to the ground, so high frequency signals  have a low resistance return path. 
This design enables the peak detectors to measure low voltages without impacting the matching network performance significantly.

The matching network needs to be co-designed with the power detectors due to the small added capacitance.
The proposed circuit schematic (with DC blocks) is shown in Fig. \ref{fig:mn_w_pds}, and the fabricated board is shown in Fig. \ref{fig:fabricated_boards}\ref{fig:mn_w_pds_pcb}.
As can be seen in Fig. \ref{fig:fabricated_boards}\ref{fig:mn_w_pds_pcb} in the physical design of the matching network, both LC sections have a microstrip inductor and a lumped capacitor. 
Microstrip inductors are used due to their high-Q and the ease of matching between different unit cells.

Normally, an ideal series component would not have a diagonalizable transmission matrix.
However, due to the parasitic capacitance of the inductors each unit cell consists of a mix of shunt and series components. 
This is also verified with the co-simulation of the design.

The co-simulation consists of two parts: EM simulation of the PCB layout, and the lumped models of the discrete devices provided by the manufacturers. 
The main assumption in the proposed theory is that there are repeating structures in the circuit that can be leveraged to extract more information. 
In the EM simulation, this assumption can be violated by how meshes are constructed. 
This can affect the measurement procedure negatively, but it can be resolved by simulating a single unit cell and measurement port.
Then, unit cells, measurement ports, discrete components and the rest of the layout are connected in the circuit simulation.
This guarantees that each unit cell and measurement port have the same response.

The design is simulated with 193 loads regularly distributed across the Smith Chart used as test loads. 
Test loads have VSWR under 10:1 ($S_{11} < -2$ dB).
Fig. \ref{fig:sim_smith_chart} shows the test load locations and the measurement results for each load. 
It is clear from Fig. \ref{fig:sim_smith_chart} that the proposed method measures $S_{11}$ with high accuracy. Simulations show that 10:1 VSWR requires \SI{19}{dB} and 3:1 VSWR requires \SI{12}{dB} of dynamic range from the power detectors. This dynamic range requirement is expected to be lower in the real hardware experiment due to additional losses resulting from the experiment.

\section{Hardware Experiment} \label{sec:experiment}
To show the feasibility of the proposed method, two hardware experiments are conducted. 
In the first experiment, it is verified that the proposed method does not degrade matching network performance.
In the second experiment, it is verified that the proposed method can measure loads with high accuracy.

\subsection{Impact on Hardware Performance}

One board with (Fig. \ref{fig:fabricated_boards}\ref{fig:mn_w_pds_pcb}) and another board without power detectors (Fig. \ref{fig:fabricated_boards}\ref{fig:mn_wo_pds_pcb}) are fabricated.
Then, two boards are compared to see whether the proposed technique degrades the performance of the matching network.
TRL calibration is used to deembed the SMA connectors from the VNA measurements.
In TRL calibration, impedances are normalized to the characteristic impedance of the transmission line ($Z_0$). Therefore, to measure the real values of the loads, $Z_0$ of the transmission line needs to be known.
A high precision \SI{50}{\ohm} load is included with the TRL calibration kit. 
Using this included known impedance, $Z_0$ is found to be \SI{46}{\ohm}, which also agrees with the EM simulation.
\begin{figure}[t!]
    \begin{minipage} {\linewidth}
        \centering
        \includegraphics[width=\textwidth]{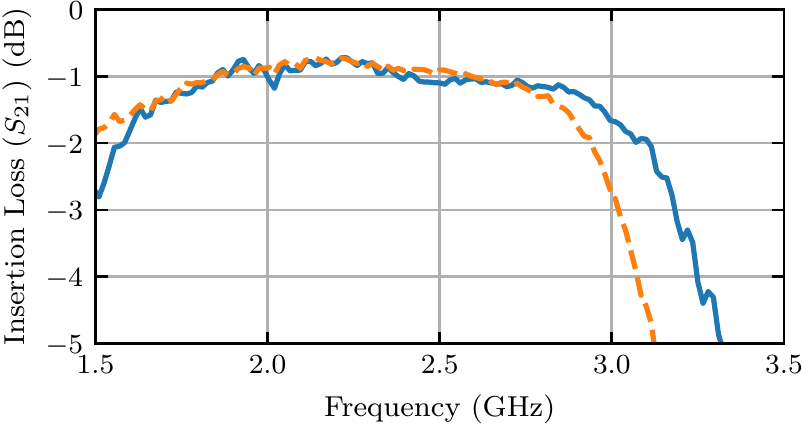}
        \Scaption{Insertion Loss}
        \label{fig:insertion_loss_comparison}
        \vspace{3mm}
    \end{minipage}
    \begin{minipage} {\linewidth}
        \centering
        \includegraphics[width=\textwidth]{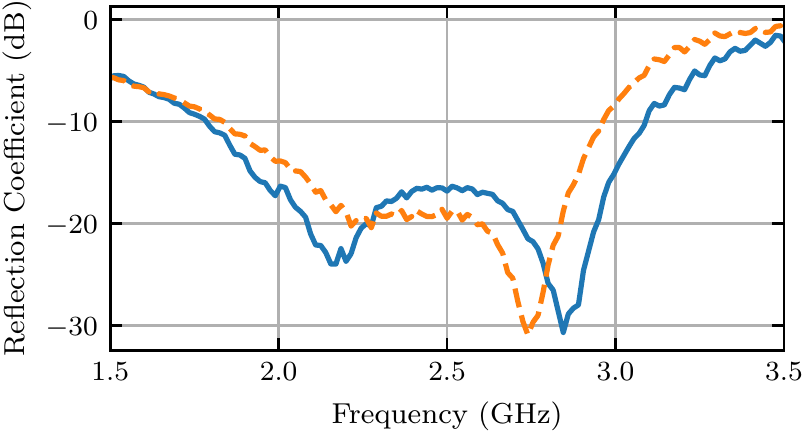}
        \Scaption{Reflection Coefficient}
        \label{fig:matching_comparison}
        \vspace{3mm}
    \end{minipage}
    \caption{Matching Network Performance with ({\color{C0} \full}) and without ({\color{C1} \dashed}) Power Detectors}
    \label{fig:board_comparison}
\end{figure}
\begin{figure} [b!]
    \centering
    \includegraphics[width=0.5\linewidth]{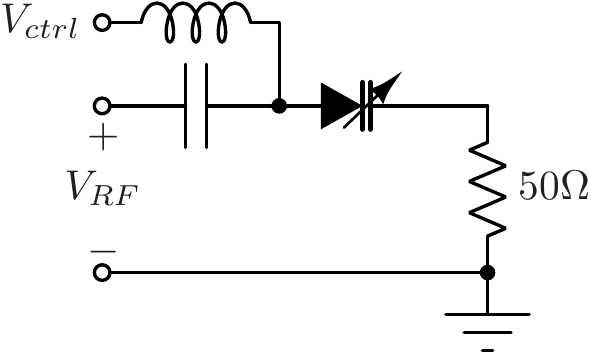}
    \caption{Adjustable Load Schematic}
    \label{fig:adj_load_schematic}
\end{figure}
When the matching network is co-designed with the proposed method, adding the power detectors does not degrade insertion loss (Fig. \ref{fig:board_comparison}\ref{fig:insertion_loss_comparison}) and impedance matching performance (Fig. \ref{fig:board_comparison}\ref{fig:matching_comparison}) of the matching network.
These results also agree with the EM simulation. 
Overall, both boards transform $50\Omega$ to $5\Omega$ with central frequency of 2.5GHz.
The reflection coefficients characteristic agrees with the EM simulation and the two minimums observed verifies that the two section matching works as expected.

\begin{figure}[t!]
    \begin{minipage} {\linewidth}
        \centering
        \includegraphics[width=\textwidth]{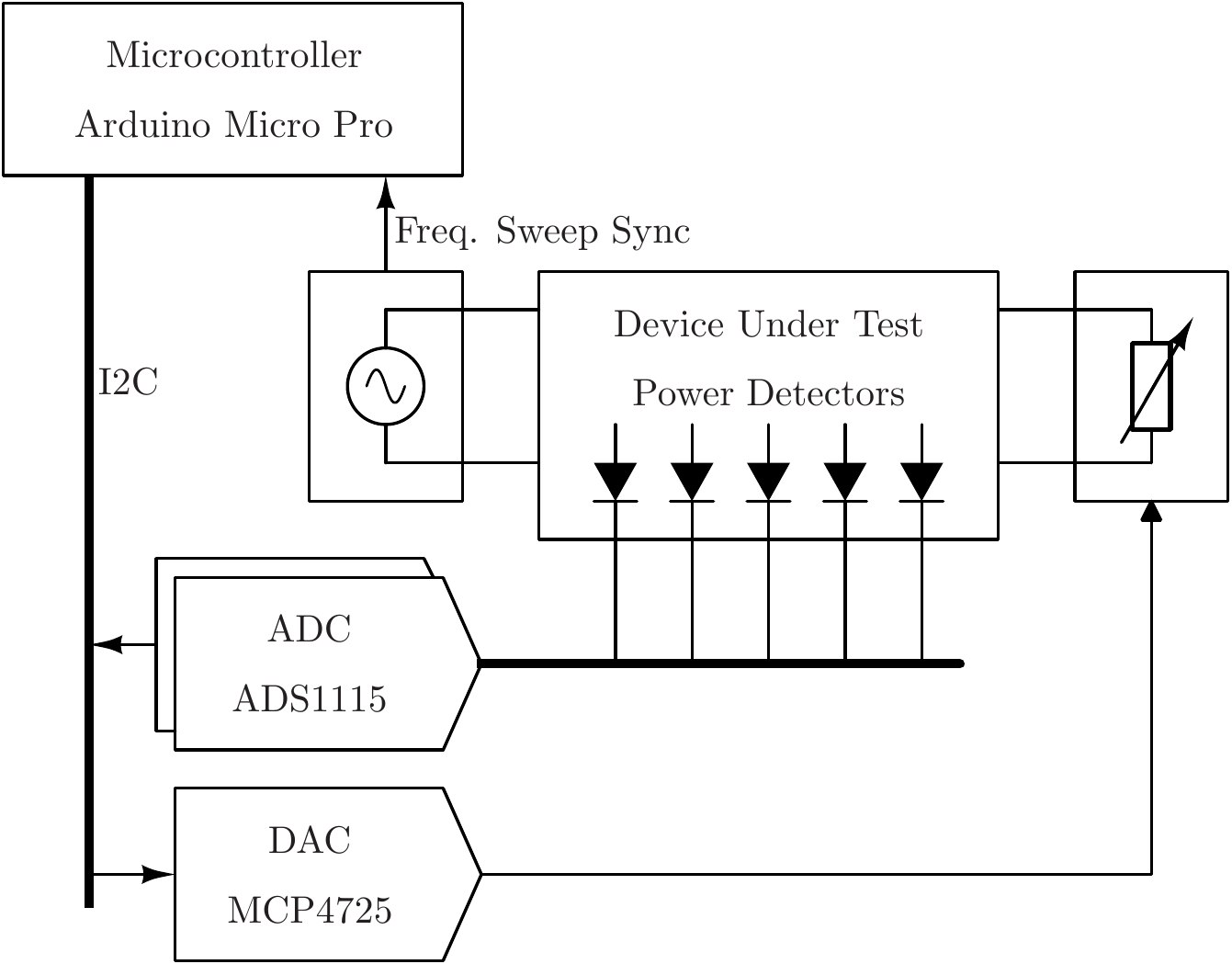}
        \Scaption{Experiment Block Diagram}
        \label{fig:experiment_block_diagram}
        \vspace{3mm}
    \end{minipage}
    \begin{minipage} {\linewidth}
        \centering
        \includegraphics[width=\textwidth]{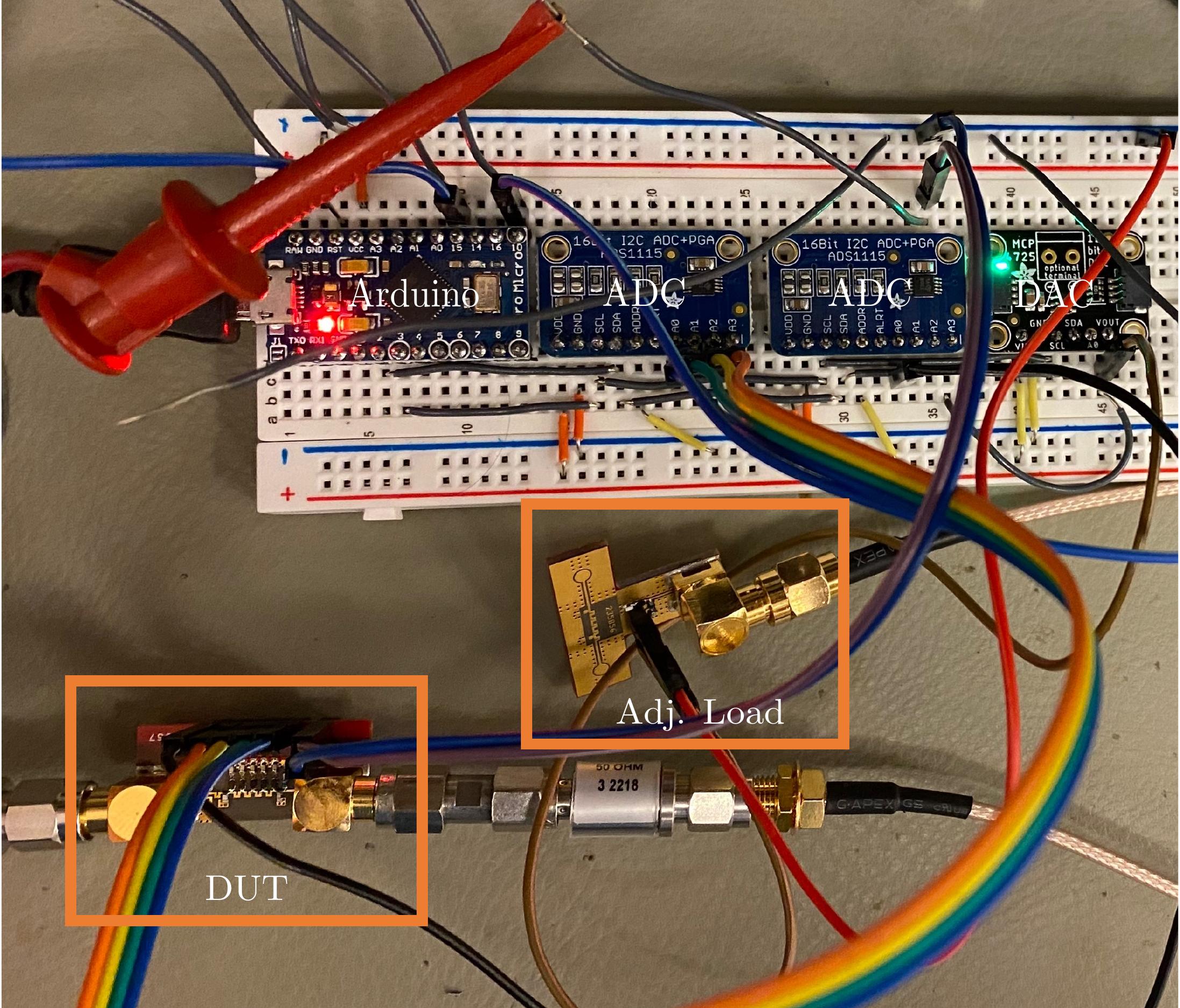}
        \Scaption{Experiment Setup}
        \label{fig:experimet_setup}
    \end{minipage}
    \caption{Hardware Experiment}
    \label{fig:hardware_experiment}
\end{figure}

\subsection{Measurement Accuracy}

An adjustable load across a wide range of frequencies is measured with a VNA (Agilent E8361A).
As shown in Fig. \ref{fig:adj_load_schematic}, the adjustable load is constructed with a diode (BAT15-02LRH) and a fixed resistor. 
When the diode is completely off ($V_{ctrl}=0$), it will act as a high impedance load looking into the RF input ($V_{RF}$). 
As the control voltage is increased ($V_{ctrl}$), the diode will become a short circuit and the fixed resistance (\SI{50}{\ohm}) will dominate the input impedance. 
There will be a rotation around the Smith Chart due to the transmission line between the adjustable load and the rest of the system. This rotation will also change with frequency.
When these two effects are combined, a wide range of loads will be created and sweep across the Smith Chart.
$V_{ctrl}$, will control how far away the load is from the center of the Smith Chart, and frequency will control the angle of the load.

\begin{figure}[tb!]
    \begin{minipage} {\linewidth}
        \centering
        \includegraphics[width=\textwidth]{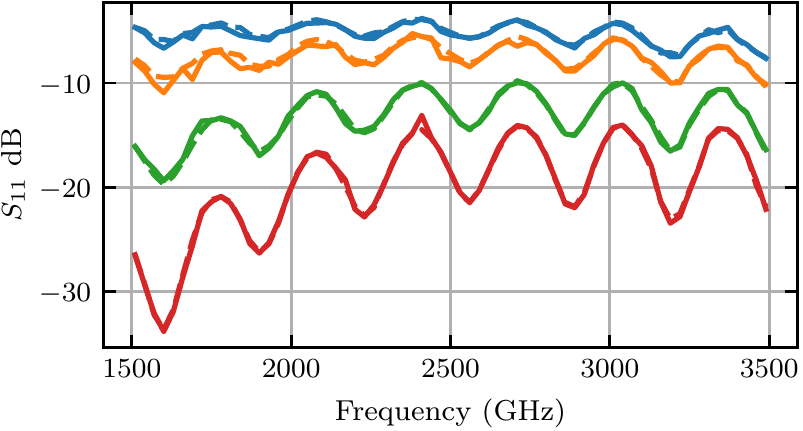}
        \Scaption{${S_{11}}$ dB}
        \label{fig:dbvsfreq}
        \vspace{3mm}
    \end{minipage}
    \begin{minipage} {\linewidth}
        \centering
        \includegraphics[width=\textwidth]{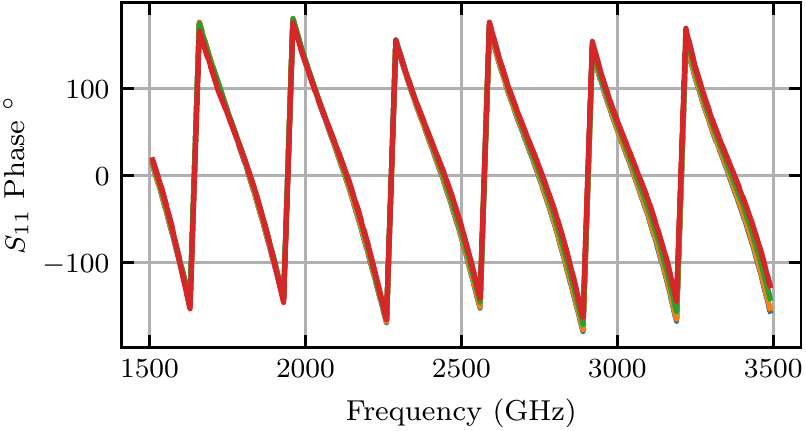}
        \Scaption{Phase of ${S_{11}}$ in degrees}
        \label{fig:phasevsfreq}
        \vspace{3mm}
    \end{minipage}
    \caption{$S_{11}$ of various test loads (Dashed) vs Measured (Solid) vs Frequency}
    \label{fig:experiment_results}
\end{figure}
 \begin{figure} [b!]
    \centering
    \includegraphics[width=\linewidth]{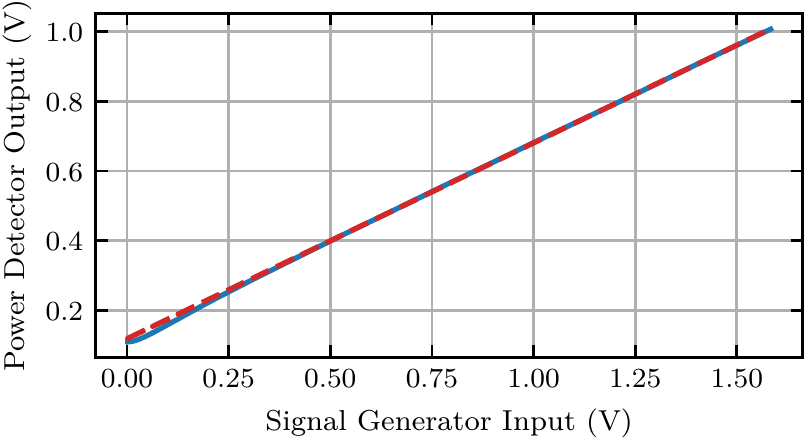}
    \caption{Power Detector Output Voltage ({\color{C0} \full}), Fitted Line ({\color{C1} $\dashed$})}
    \label{fig:power_detector_linear_model}
\end{figure}

\begin{figure}[tb!]
    \begin{minipage} {\linewidth}
        \centering
        \includegraphics[width=\textwidth]{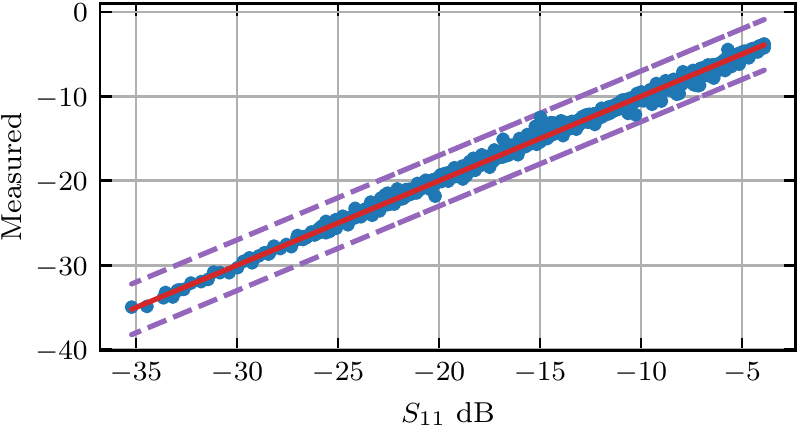}
        \Scaption{${S_{11}}$ dB, 3 dB error line ({\color{C4} \dashed}) }
        \label{fig:dbcomp}
        \vspace{3mm}
    \end{minipage}
    \begin{minipage} {\linewidth}
        \centering
        \includegraphics[width=\textwidth]{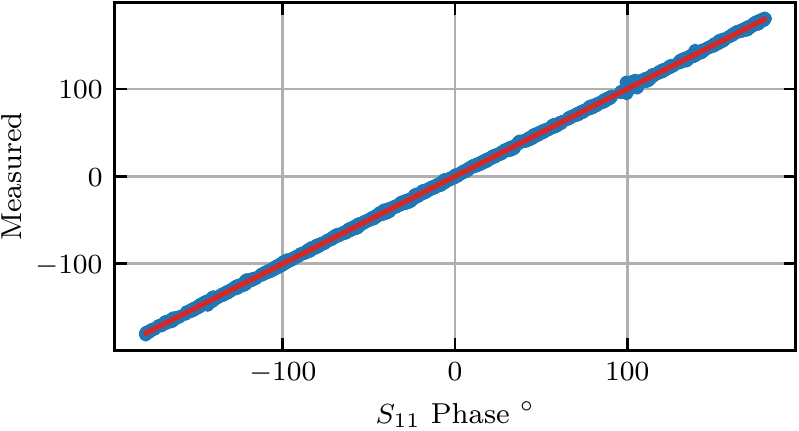}
        \Scaption{Phase of ${S_{11}}$ in degree}
        \label{fig:phase_comp}
        \vspace{3mm}
    \end{minipage}
    \caption{$S_{11}$ of All Test Loads ({\color{C3} \full}) vs Measurement ({\color{C0} $\bullet$})}
    \label{fig:experiment_results_detailed}
\end{figure}
 \begin{figure} [b!]
    \centering
    \includegraphics[width=0.75\linewidth]{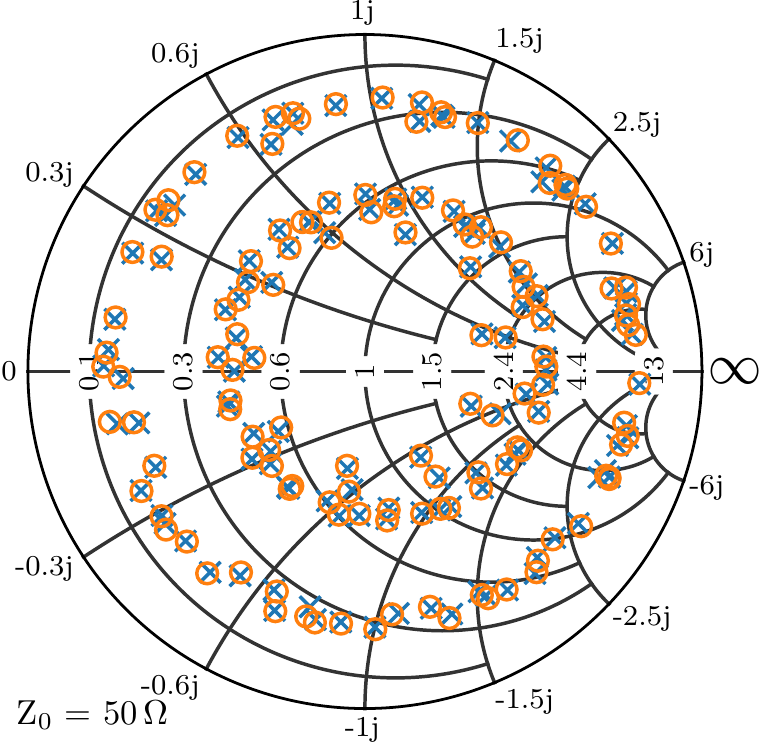}
    \caption{Design Experiment Results - Test Loads ({\color{C1} $\circ$}), Measured Values ({\color{C2} $\cross$})}
    \label{fig:smith_chart}
\end{figure}

A signal generator is connected to the board. The frequency sweep and sweep run outputs of the signal generator are connected to a microcontroller (Arduino) board.
These outputs will facilitate the frequency synchronization between the signal generator and measurements.
The control voltage for the adjustable load is generated by a DAC (MCP4725).
The outputs of the peak detectors are connected to ADC (ADS1115).
The entire process is coordinated by the microcontroller connected to a computer over USB. 
The experiment diagram and a picture of the setup are shown in Fig. \ref{fig:hardware_experiment}.

Power detectors are characterized with a power sweep from \SI{-30}{dBm} to \SI{17}{dBm} with \SI{1}{dB} steps. Power sweep values and corresponding DC output from a power detector are plotted against each other in Fig. \ref{fig:power_detector_linear_model}. 
A line is fitted in this graph to determine the DC offset and the dynamic range of the power detectors. 
The lower and upper limits of the detector outputs are determined using this model. 
The output voltage is within 2\% of the linear trend from \SI{0.28}{V} to \SI{1.58}{V}, corresponding to a \SI{15}{dB} dynamic range. 

By using adjustable load and frequency sweep, 1407 test data points (excluding calibration data points) are obtained. 
The performance of the proposed method is evaluated across frequency, different reflection coefficients and phases. 
The design center frequency is \SI{2.5}{\giga\hertz}, and the method can reliably work from \SI{1.5}{\giga\hertz} to \SI{3.5}{\giga\hertz} (Fig. \ref{fig:experiment_results}), which corresponds to 80\% fractional bandwidth. 
This means, the proposed method can easily work with high bandwidth antennas. 
The test loads require at most \SI{13.5}{dB} of dynamic range in a single measurement, which is consistent with the simulation results.

Fig. \ref{fig:experiment_results} shows the system performance across frequency. 
Fig. \ref{fig:smith_chart} shows the comparison between test loads and measured values in the Smith Chart.
Not all data points are included in these two figures to enable readability. The included data points are selected at random and to cover of a wide range of the Smith Chart.
Fig. \ref{fig:experiment_results_detailed} shows the magnitude and phase of the measured $S_{11}$ with respect to the baseline (measured by the VNA) for all loads.
Fig. \ref{fig:experiment_results}-\ref{fig:experiment_results_detailed}, do not include any calibration loads  since they have zero measurement errors by definition. 

These results show that, proposed design and method can measure a wide variety of reflection coefficients with high accuracy.
Another observation is the max. dB error increases when low magnitudes are measured. 
This is expected because, as the magnitude drops, even very small measurement errors in magnitude would result in significant errors in dB domain.
The highest dB error is 1.6 dB for loads that have $S_{11} > \SI{-10}{\dB}$.
Overall, the system can measure $S_{11}$ phase very reliably. 
The error summary is given in Table-\ref{tab:error_summary}.

The biggest constraint that limits the accuracy of the system is the linearity of the power detectors. It has been observed that, accuracy can be changed significantly if input power is above or below certain values. Therefore, input power should be carefully selected so that power detectors are working in the linear region. 

\begin{table}[t!]
\caption{Error Summary Table}
\begin{center}
\begin{tabular}{c|c|c|}
\cline{2-3}
\textbf{}&\multicolumn{2}{c|}{\textbf{All Loads}} \\
\cline{2-3} 
\textbf{} & \textbf{\textit{$S_{11}$ Mag. Error (dB)}}& \textbf{\textit{ $S_{11}$ Phase Error ($^\circ$) }} \\
\hline
\multicolumn{1}{|c|}{Max}               & 2.5 dB & 7.9$^\circ$ \\
\multicolumn{1}{|c|}{Avg$^{\mathrm{*}}$}& 0.23 dB & 0.53$^\circ$ \\
\hline
\textbf{}&\multicolumn{2}{c|}{\textbf{Loads $S_{11} > \SI{-10}{\dB}$}} \\
\cline{2-3} 
\textbf{} & \textbf{\textit{$S_{11}$ Mag. Error (dB)}}& \textbf{\textit{ $S_{11}$ Phase Error ($^\circ$) }} \\
\hline
\multicolumn{1}{|c|}{Max}               & 1.57 dB & 4.3$^\circ$ \\
\multicolumn{1}{|c|}{Avg$^{\mathrm{*}}$}& 0.26 dB & 0.5$^\circ$ \\
\hline
\multicolumn{3}{l}{$^{\mathrm{*}}$ Calibration loads are not included.}
\end{tabular}
\label{tab:error_summary}
\end{center}
\end{table}

\section*{Conclusion}
In this work, a new multi-port architecture that uses periodic structures is proposed. 
This new structure results in a new model for multi-ports, which greatly simplifies equations.
This simplification results in closed-form calibration and measurements for the multi-ports. 
Also, calibration can be done by just using three loads, which was not possible in generic SPR architectures.
The proposed theory is verified with Monte-Carlo, circuit, and EM simulations.
Experimental results show that the proposed method gives excellent accuracy for a wide range of frequencies and loads.

\section*{Future Work}
This work is focused on measurement systems that consist of repetition of the same block, (blocks that have the same parameters with equal values). 
This can be generalized with blocks that can be described by the different parameters with known relationships.
A prior knowledge between parameters can be used to deduce connections between impedances and voltage magnitude measurements. 
This could result in a richer family of BIST/measurement system configurations, e.g., mismatch detectors embedded in filters, matching networks etc. 

\section*{Appendix}
 \begin{figure} [h]
    \centering
    \includegraphics[width=0.25\textwidth]{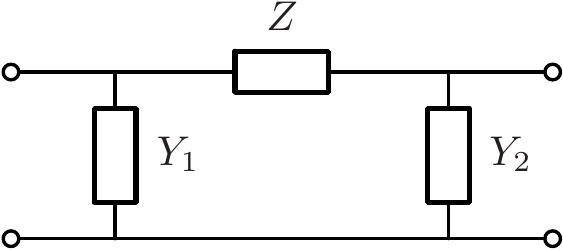}
    \caption{$\pi$-network representation of a reciprocal two-port linear network }
    \label{fig:pi_network}
\end{figure}
Any reciprocal linear network can be modeled as the circuit given in Fig. \ref{fig:pi_network}. $Z$, $Y_1$ and $Y_2$ values can be calculated using Y-parameters.
Given $Z$, $Y_1$ and $Y_2$, the transmission matrix of this unit cell is:
\begin{align}
    \matr{T} = \left[\begin{matrix}Y_{2} Z + 1 & Z\\Y_{1} + Y_{1} Y_{2} Z + Y_{2} & Y_{1} Z + 1\end{matrix}\right].
\end{align}
The eigenvalues of this matrix are:
\begin{align}
    \lambda_{1,2} = 1 + \frac{Z (Y_{1} + Y_{2}) \pm \sqrt{Z \left(Y_{1} + Y_{2}\right) \left(Z(Y_{1} + Y_{2}) + 4\right)}}{2} .
\end{align}
Therefore, if the following conditions are met:
\begin{itemize}
  \item There is a series element in the equivalent circuit ($Z \neq 0$) 
  \item There is at least one shunt element in the equivalent circuit, and the sum of the admittance of the shunt elements is not zero ($Y_1 + Y_2 \neq 0 $)
  \item Multiplication of the series element impedance and the sum of the admittance of the shunt elements satisfies the following $Z(Y_1 + Y_2) + 4 \neq 0$
\end{itemize}
The transmission matrix of the unit cell will be diagonalizable.

{\tiny
\bibliographystyle{IEEEtran}
\bibliography{IEEEabrv, references}
}

\end{document}